\documentclass[11pt,a4paper]{article}
\usepackage{amssymb}
\usepackage{times,cite}
\newcommand{\be}{\begin{eqnarray}}
\newcommand{\ee}{\end{eqnarray}}
\newcommand{\bez}{\begin{eqnarray*}}
\newcommand{\eez}{\end{eqnarray*}}
\newcommand{\pa}{\partial}
\newcommand{\la}{\lambda}
\newcommand{\ci}{\circ}
\newcommand{\tr}{\mathrm{tr}}

\newcommand{\A}{\mathbb{A}}
\newcommand{\N}{\mathbb{N}}
\newcommand{\cA}{\mathcal{A}}
\newcommand{\R}{\mathcal{R}}

\newcommand{\res}{\mathrm{res}}
\newcommand{\fA}{\mathfrak{A}}

\title{\bf Weakly nonassociative algebras, Riccati and 
       KP hierarchies}

\author{Aristophanes Dimakis$^a$ and Folkert M\"uller-Hoissen$^b$ \\
 $^a$ Department of Financial and Management Engineering, 
 University of the Aegean \\
 31 Fostini Str., GR-82100 Chios, Greece \\
 $^b$ Max-Planck-Institute for Dynamics and Self-Organization \\
 Bunsenstrasse 10, D-37073 G\"ottingen, Germany \\
 E-mails: dimakis@aegean.gr, folkert.mueller-hoissen@ds.mpg.de }

\date{}

\begin{document}

\renewcommand{\theequation} {\arabic{section}.\arabic{equation}}

\newtheorem{theorem}{Theorem}
\newtheorem{lemma}{Lemma}
\newtheorem{proposition}{Proposition}
\newtheorem{definition}{Definition}
\newtheorem{corollary}{Corollary}

\maketitle

\vspace{-1.cm}

\begin{abstract}
It has recently been observed that certain nonassociative algebras 
(called `weakly nonassociative', WNA) determine, via a universal 
hierarchy of ordinary differential equations, 
solutions of the KP hierarchy with dependent variable in an associative 
subalgebra (the middle nucleus). We recall central results and consider 
a class of WNA algebras for which the hierarchy of ODEs reduces to 
a matrix Riccati hierarchy, which can be easily solved.
The resulting solutions of a matrix KP hierarchy determine, under 
a `rank one condition', solutions of the scalar KP hierarchy. We 
extend these results to the discrete KP hierarchy. Moreover, we build 
a bridge from the WNA framework to the Gelfand-Dickey formulation of 
the KP hierarchy.
\end{abstract}

\section{Introduction}
\label{section:intro}
\setcounter{equation}{0}
The Kadomtsev-Petviashvili (KP) equation is an extension of 
the famous Korteweg-deVries (KdV) equation to 2+1 dimensions. It first 
appeared in a stability analysis of KdV solitons \cite{Kado+Petv70,Infe+Rowl00}. 
In particular, it describes nonlinear fluid surface waves in a certain 
approximation and explains to some extent the formation of network patterns 
formed by line wave segments on a water surface \cite{Infe+Rowl00}. 
It is `integrable' in several respects, in particular in the sense 
of the inverse scattering method. Various remarkable properties 
have been discovered that allow to access (subsets of) its solutions 
in different ways, see in particular \cite{Matv+Sall91,MJD00,Dick03}. 
Apart from its direct relevance in physics, the KP equation and its hierarchy 
(see \cite{Dick03,Kupe00}, for example) is deeply related to  
the theory of Riemann surfaces (Riemann-Schottky problem, see
\cite{Buch+Kric06} for a review). Some time ago, this stimulated 
discussions concerning the role of KP in string theory (see \cite{AGR87b,Mula88,Gilb88,Morozov92}, for example). 
Later the Gelfand-Dickey hierarchies, of which the KdV hierarchy is 
the simplest and which are reductions of the KP hierarchy, made 
their appearance in matrix models, first in a model of two-dimensional 
quantum gravity (see \cite{Morozov94,Mula94matrix} and references therein). 
This led to important developments in algebraic geometry 
(see \cite{Witt93_topmeth}, for example). Of course, what we 
mentioned here by far does not exhaust what is known about KP and 
there is probably even much more in the world of mathematics and physics 
linked to the KP equation and its descendants that still waits 
to be uncovered. 

In fact, an apparently completely different appearance of the KP hierarchy 
has been observed in \cite{DMH06nahier}. 
On a freely generated `weakly nonassociative' (WNA) algebra 
(see section~\ref{section:na}) there is a family of commuting 
derivations\footnote{Families of commuting derivations 
on certain algebras also appeared in \cite{IKZ04,Ihara05}, for example. 
In fact, the ideas underlying the work in \cite{DMH06nahier} grew 
out of our work in \cite{DMH05KPalgebra} which has some algebraic 
overlap with \cite{IKZ04}. } 
that satisfy identities which are in correspondence with the equations 
of the KP hierarchy (with dependent variable in a noncommutative 
associative subalgebra). As a consequence, there is a hierarchy of 
ordinary differential equations (ODEs) on this WNA algebra that 
implies the KP hierarchy. More generally, this holds for \emph{any} 
WNA algebra. In this way WNA algebras determine classes of solutions 
of the KP hierarchy.

In section~\ref{section:na} we recall central results of \cite{DMH06nahier} 
and present a new result in proposition~\ref{prop:naBT}. Section~\ref{section:matrix} 
applies the WNA approach to derive a matrix Riccati\footnote{Besides 
their appearance in control and systems theory, matrix Riccati equations 
(see \cite{Reid72,Zakh73,AFIJ03}, for example) frequently showed up in the 
context of integrable systems, see in particular
\cite{Chau83,HSS84,Taka85,Comm+Robe86,DNS89,Grun+Levi97,FGRSZ99,Haze91,Haak96,FMP98,Taka89,FMPZ00,Zeli03}. }
hierarchy, the solutions of which are solutions of the corresponding 
matrix KP hierarchy (which under certain conditions determines solutions of 
the scalar KP hierarchy).
In section~\ref{section:DKP} we extend these results to the 
discrete KP hierarchy \cite{Adle+vanM99,Dick99,Hain+Ilie00,Feli+Onga01,DMH06func}.  
Furthermore, in section~\ref{section:GD} we show how the Gelfand-Dickey 
formulation \cite{Dick03} of the KP hierarchy (with dependent variable in 
any associative algebra) emerges in the WNA framework. 
Section~\ref{section:concl} contains some conclusions.

\section{Nonassociativity and KP} 
\label{section:na}
\setcounter{equation}{0}
In \cite{DMH06nahier} we called an algebra $(\A,\circ)$ (over a commutative ring) 
\emph{weakly nonassociative (WNA)} if 
\be
    (a,b \circ c,d) = 0  \qquad \quad  \forall a,b,c,d \in \A  \, ,  \label{WNA}
\ee
where $(a,b,c) := (a \circ b) \circ c - a \circ (b \circ c)$ is the associator in $\A$. 
The \emph{middle nucleus} of $\A$ (see e.g. \cite{Pflug90}), 
\be
    \A' := \{ b \in \A  \; | \; (a,b,c) =0 \; \; \forall a,c \in \A \} \, , 
\ee 
is an \emph{associative} subalgebra and a two-sided ideal. We fix 
$f \in \A$, $f \not\in \A'$, and define $a \circ_1 b := a \circ b$, 
\be
    a \circ_{n+1} b := a \circ (f \circ_n b) - (a \circ f) \circ_n b  \, ,
        \qquad \quad  n=1,2, \ldots \; .    \label{circ_n}
\ee
As a consequence of (\ref{WNA}), these products only depend on the equivalence 
class $[f]$ of $f$ in $\A/\A'$.
The subalgebra $\A(f)$, generated by $f$ in the WNA algebra $\A$, 
is called \emph{$\delta$-compatible} if, for each $n \in \N$, 
\be
        \delta_n(f) := f \circ_n f    \label{delta_n}
\ee
extends to a \emph{derivation} of $\A(f)$. 
In the following we recall some results from \cite{DMH06nahier}. 

\begin{theorem}
\label{theorem:delta-KP}
Let $\A(f)$ be $\delta$-compatible. The derivations $\delta_n$ 
commute on $\A(f)$ and satisfy identities that are in correspondence via 
$\delta_n \mapsto \pa_{t_n}$ (the partial derivative operator 
with respect to a variable $t_n$) 
with the equations of the potential Kadomtsev-Petviashvili (pKP) hierarchy 
with dependent variable in $\A'$. 
\hfill $\square$
\end{theorem}

This is a central observation in \cite{DMH06nahier} with the following 
immediate consequence.

\begin{theorem}
\label{theorem:WNA-KP}
Let $\A$ be any WNA algebra over the ring of complex functions 
of independent variables $t_1,t_2, \ldots$. If $f \in \A$ solves the  
hierarchy of ODEs\footnote{$f$ has to be differentiable, of course, which 
requires a corresponding (e.g. Banach space) structure on $\A$. 
The flows given by (\ref{na_hier}) indeed commute \cite{DMH06nahier}. 
Furthermore, (\ref{na_hier}) implies $\delta$-compatibility of the algebra 
$\A(f)$ generated by $f$ in $\A$ over $\mathbb{C}$ \cite{DMH06nahier}. }
\be
   f_{t_n} := \pa_{t_n}(f) = f \circ_n f \, , \qquad \qquad n=1,2, \ldots 
         \, ,  \label{na_hier}
\ee
then $-f_{t_1}$ lies in $\A'$ and solves the KP hierarchy 
with dependent variable in $\A'$. 
\hfill $\square$
\end{theorem}

\begin{corollary}
If there is a constant $\nu \in \A$, $\nu \not\in \A'$, with $[\nu] = [f] \in \A/\A'$, 
then, under the assumptions of theorem~\ref{theorem:WNA-KP}, 
\be
    \phi := \nu - f \in \A'    \label{phi=nu-f}
\ee
solves the potential KP (pKP) hierarchy\footnote{This functional representation 
of the potential KP hierarchy appeared in \cite{Bogd+Kono98,Bogd99}. See also 
\cite{DNS89,DMH06nahier} for equivalent formulae.}
\be
   \sum_{i,j,k=1}^3 \varepsilon_{ijk} \, \left( \la_i^{-1} (\phi_{[\la_i]} - \phi)
      + \phi \circ \phi_{[\la_i]} \right)_{[\la_k]} = 0  
          \, ,  \label{functionalpKP}
\ee
where $\varepsilon_{ijk}$ is totally antisymmetric with $\varepsilon_{123} = 1$,  
$\la_i$, $i=1,2,3$, are indeterminates, and 
$\phi_{\pm [\la]}(\mathbf{t}) := \phi(\mathbf{t} \pm [\la])$, 
where $\mathbf{t}=(t_1,t_2,\ldots)$ and $[\la] := (\la, \la^2/2, \la^3/3, \ldots)$. 
\hfill $\square$
\end{corollary}

\noindent
\emph{Remark~1.} If $C \in \A'$ is constant, then $f = \nu' - (\phi + C)$ 
with constant $\nu' := \nu + C$ satisfying $[\nu']=[\nu]=[f]$. Hence, with 
$\phi$ also $\phi + C$ is a solution of the pKP hierarchy. This can also 
be checked directly using (\ref{functionalpKP}), of course.
\hfill $\square$
\vskip.1cm

The next result will be used in section~\ref{section:DKP}. 

\begin{proposition}
\label{prop:naBT}
Suppose $f$ and $f'$ solve (\ref{na_hier}) and $[f]=[f']$ in a WNA algebra $\A$. 
The equation
\be
    f' \circ f = \alpha \, (f' - f)   \label{naBT}
\ee
is then preserved for all $\alpha \in \mathbb{C}$. 
\end{proposition} 
{\em Proof:} 
\vspace{-.6cm}
\bez
   (f' \circ f)_{t_n} 
 &=& f'_{t_n} \circ f + f' \circ f_{t_n} 
  = (f' \ci_n f') \circ f + f' \circ (f \ci_n f) \\
 &=& (f' \ci_n f') \circ f - f' \ci_n (f' \circ f) 
     + \alpha \, f' \ci_n (f' -f)  \\
 & & + f' \circ (f \ci_n f) - (f' \circ f) \ci_n f 
     + \alpha \, (f' -f) \ci_n f \\
 &=& - f' \ci_{n+1} f + f' \ci_{n+1} f + \alpha \, ( f' \ci_n f' - f \ci_n f)
  = \alpha \, (f' -f)_{t_n} \; .
\eez
In the third step we have added terms that vanish as a consequence 
of (\ref{naBT}). 
Then we used (3.11) in \cite{DMH06nahier} (together with the fact that the 
products $\ci_n$ only depend on the equivalence class $[f]=[f'] \in \A/\A'$),
and also (\ref{circ_n}), to combine pairs of terms into products of one 
degree higher.
\hfill $\square$
\vskip.1cm

\noindent
\emph{Remark~2.} In functional form, (\ref{na_hier}) can be expressed 
(e.g. with the help of results in \cite{DMH06nahier}) as
\be
     \la^{-1} ( f - f_{-[\la]} ) - f_{-[\la]} \circ f = 0  \; .
         \label{na_hier_func}
\ee
Setting $f' = f_{-[\la]}$ (which also solves (\ref{na_hier_func}) if $f$ solves it), 
this takes the form (\ref{naBT}) with $\alpha = -\la^{-1}$. 
\hfill $\square$
\vskip.1cm

In order to apply the above results, we need examples of WNA algebras. 
For our purposes, it is sufficient to recall from \cite{DMH06nahier} that 
any WNA algebra with $\mathrm{dim}(\A/\A')=1$ is isomorphic to one 
determined by the following data: \\
(1) an associative algebra $\mathcal{A}$ (e.g. any matrix algebra) \\
(2) a fixed element $g \in \mathcal{A}$ \\
(3) linear maps $\mathcal{L}, \mathcal{R} \, : \, \mathcal{A} \rightarrow \mathcal{A}$ such that 
\be  
    [\mathcal{L},\mathcal{R}] =0 \, , \qquad 
    \mathcal{L}(a \circ b) = \mathcal{L}(a) \circ b \, , \qquad 
    \mathcal{R}(a \circ b) = a \circ \mathcal{R}(b) \; .  \label{LR-rels}
\ee
Augmenting $\mathcal{A}$ with an element $f$ such that
\be
   f \circ f := g \, , \qquad f \circ a := \mathcal{L}(a) \, , \qquad
   a \circ f := \mathcal{R}(a) \, ,    \label{f-augm}
\ee
leads to a WNA algebra $\A$ with $\A' = \mathcal{A}$, provided that the 
following condition holds,
\be
    \exists \, a,b \in \mathcal{A} \, : \, 
    \mathcal{R}(a) \circ b \neq a \circ \mathcal{L}(b) \; . 
             \label{na_condition}
\ee
This guarantees that the augmented algebra is \emph{not} associative. 
Particular examples of $\mathcal{L}$ and $\mathcal{R}$ are given 
by multiplication from left, respectively right, by fixed elements 
of $\mathcal{A}$ (see also the next section).

\section{A class of WNA algebras and a matrix Riccati hierarchy}
\label{section:matrix}
\setcounter{equation}{0}
Let $\mathcal{M}(M,N)$ be the vector space of complex 
$M \times N$ matrices, depending smoothly on independent real variables 
$t_1,t_2,\ldots$, and let $S,L,R,Q$ be constant matrices of 
dimensions $M \times N$, $M \times M$, $N \times N$ and $N \times M$, 
respectively. 
Augmenting with a constant element $\nu$ and setting\footnote{Using (\ref{phi=nu-f}), in terms of $f$ this yields relations of the form 
(\ref{f-augm}).}
\be
    \nu \circ \nu = -S \, , \qquad \nu \circ A = L A \, , \qquad
    A \circ \nu = -A R \, , \qquad A \circ B = A Q B \, , 
\ee
for all $A,B \in \mathcal{M}(M,N)$, we obtain a WNA algebra $(\A,\circ)$. 
The condition (\ref{na_condition}) requires 
\be
    R Q \neq Q L \; .      \label{matrix_na_condition}
\ee 
 For the products $\ci_n$, $n>1$, we have the following result. 

\begin{proposition}
\be
    \nu \ci_n \nu = - S_n \, , \qquad
    \nu \ci_n A = L_n A \, , \qquad
    A \ci_n \nu = - A R_n \, , \qquad
    A \ci_n B = A Q_n B \, ,     \label{circ_n-recursion}
\ee
where
\be
    \left(\begin{array}{cc} R_n & Q_n \\ S_n & L_n \end{array}\right) = H^n 
        \label{H_relation}
    \qquad
\mbox{with} \quad
    H := \left(\begin{array}{cc} R & Q \\ S & L \end{array}\right) \; .
\ee
\end{proposition}
{\em Proof:} Using the definition (\ref{circ_n}), one proves by induction that
\bez
    S_{n+1} = L S_n + S R_n  \, ,  \quad & L_{n+1} = L L_n + S Q_n \, , \quad 
    R_{n+1} = Q S_n + R R_n  \, ,  \quad & Q_{n+1} = Q L_n + R Q_n \, , 
\eez
for $n=1,2,\ldots$, where $S_1 = S$, $L_1 = L$, $R_1 = R$, $Q_1 = Q$. 
This can be written as
\bez
    \left(\begin{array}{cc} R_{n+1} & Q_{n+1} \\ S_{n+1} & L_{n+1}
          \end{array}\right) 
   = H \, \left(\begin{array}{cc} R_n & Q_n \\ S_n & L_n \end{array}\right) \, ,
\eez
which implies (\ref{H_relation}). 
\hfill $\square$
\vskip.1cm

Using (\ref{phi=nu-f}) and (\ref{circ_n-recursion}) in (\ref{na_hier}), 
leads to the matrix Riccati equations\footnote{\label{foot:mRic_func} 
The corresponding functional form is 
$\la^{-1}(\phi_{[\la]}-\phi) + \phi Q \phi_{[\la]} = S + L \phi_{[\la]} - \phi R$, 
which is easily seen to imply (\ref{functionalpKP}), see also \cite{DMH06Burgers}. 
The appendix provides a FORM program \cite{Heck00,Verm02} 
which independently verifies that any solution of (\ref{matrix_Riccati}), 
reduced to $n=1,2,3$, indeed solves the matrix pKP equation in 
$(\mathcal{M}(M,N),\circ)$. }
\be
    \phi_{t_n} = S_n + L_n \phi - \phi R_n - \phi Q_n \phi \, ,
    \qquad n=1,2, \ldots \; .       \label{matrix_Riccati}
\ee
Solutions of (\ref{matrix_Riccati}) are obtained in a well-known way 
(see \cite{AFIJ03,Taka89}, for example) via
\be
     \phi = Y X^{-1}
\ee
from the linear system
\be
    Z_{t_n} = H^n Z \, , \qquad 
    Z = \left(\begin{array}{c} X \\ Y \end{array}\right) 
       \label{Z-linsys}
\ee
with an $N \times N$ matrix $X$ and an $M \times N$ matrix $Y$, provided 
$X$ is invertible. This system is solved by
\be
    Z(\mathbf{t}) = e^{\xi(H)} Z_0 \qquad \mbox{where} \quad
    \xi(H) := \sum_{n \geq 1} t_n H^n \; .  \label{Z-sol}
\ee

If $Q$ has rank 1, then
\be
   \varphi := \tr(Q \phi)    \label{varphi}
\ee
defines a homomorphism from $(\mathcal{M}(M,N),\circ)$ into the scalars (with 
the ordinary product of functions). Hence, if $\phi$ solves the pKP hierarchy 
in $(\mathcal{M}(M,N),\circ)$, then $\varphi$ solves the scalar 
pKP hierarchy.\footnote{For related results and other perspectives on the 
rank one condition, see \cite{Gekh+Kasm06} and the references cited there. 
The idea to look for (simple) solutions of matrix and more generally operator
versions of an `integrable' equation, and to generate from it (complicated)
solutions of the scalar equation by use of a suitable map, already appeared 
in \cite{March88} (see also  
\cite{Aden+Carl96,Carl+Schi99,Blohm00,Carl+Schi00,Han+Li01,Huan03,Schie02,Schie05}). }
More generally, if $Q = V U^T$ with $V, U$ of 
dimensions $N \times r$, respectively $M \times r$, 
then $U^T \phi V$ solves the $r \times r$-matrix KP hierarchy.
\vskip.1cm

$GL(N+M,\mathbb{C})$ acts on the space of all $(N+M) \times (N+M)$ 
matrices $H$ by similarity transformations. 
In a given orbit this allows to choose for $H$ some 
`normal form', for which we can evaluate (\ref{Z-sol}) and then 
elaborate the effect of $GL(N+M,\mathbb{C})$ transformations (see also 
remark~3 below) on the corresponding solution of the pKP hierarchy, 
with the respective $Q$ given by the normal form of $H$. 
By a similarity transformation we can always achieve that $Q=0$ 
and the problem of solving the pKP hierarchy 
(with some non-zero $Q$) can thus in principle be reduced to solving 
its linear part. Alternatively, we can always achieve that $S=0$ and 
the next two examples take this route. 
\vskip.1cm

\noindent
{\bf Example 1.} If $S=0$, we can in general not achieve that also $Q=0$. 
In fact, the matrices 
\be 
   H = \left( \begin{array}{cc} R & Q \\ 0 & L \end{array} \right) 
    \quad \mbox{and} \quad
   H_0 := \left( \begin{array}{cc} R & 0 \\ 0 & L \end{array} \right) 
\ee
are similar (i.e. related by a similarity transformation) if and only 
if the matrix equation $Q = R K - K L$ has an $N \times M$ matrix 
solution $K$ \cite{Roth52,Flan+Wimm77,Olsh92,Bhat+Rose97,Schw82}, 
and then 
\be
  H = \mathcal{T} \, H_0 \, \mathcal{T}^{-1} \, , \qquad 
  \mathcal{T} = \left( \begin{array}{cc} I_N & -K \\ 0 & I_M \end{array} \right)
     \; .    \label{H-H0-ex1}
\ee
It follows that
\be
  H^n = \mathcal{T} \, H_0^n \, \mathcal{T}^{-1}
      = \left(\begin{array}{cc} R^n & R^n K - K L^n \\ 
                                0 & L^n \end{array} \right) 
\ee
and thus
\be
   e^{\xi(H)} 
 = \left(\begin{array}{cc} e^{\xi(R)} & e^{\xi(R)} K - K e^{\xi(L)}
    \\ 0 & e^{\xi(L)} \end{array}\right)  \; .
\ee
If (\ref{matrix_na_condition}) holds, we obtain the following solution of 
the matrix pKP hierarchy in $(\mathcal{M}(M,N),\circ)$, 
\be
    \phi
 = e^{\xi(L)} \phi_0 \, (I_N + K \phi_0 - e^{-\xi(R)} K e^{\xi(L)} \phi_0)^{-1}
   e^{-\xi(R)} \, ,
\ee
where $\phi_0 = Y_0 X_0^{-1}$. This in turn leads to 
\be
     \varphi 
 &=& \tr\Big( e^{-\xi(R)} (RK-KL) e^{\xi(L)} \phi_0 \, 
        (I_N + K \phi_0 - e^{-\xi(R)} K e^{\xi(L)} \phi_0)^{-1} \Big) 
     \nonumber \\
 &=& \tr\Big( \log(I_N + K \phi_0 - e^{-\xi(R)} K e^{\xi(L)} \phi_0) \Big)_{t_1}
     \nonumber \\
 &=& (\log \tau)_{t_1}  \, , \qquad \quad
     \tau := \det(I_N + K \phi_0 - e^{-\xi(R)} K e^{\xi(L)} \phi_0) \; .
\ee
If $\mathrm{rank}(Q)=1$, then $\varphi$ solves the scalar pKP hierarchy. 
Besides (\ref{matrix_na_condition}) and this rank condition, further 
conditions will have to be imposed on the (otherwise arbitrary) matrices 
$R,K,L$ and $\phi_0$ to achieve that $\varphi$ is a \emph{real} and 
\emph{regular} solution. 
See \cite{DMH06CJP}, and references cited there, 
for classes of solutions obtained from an equivalent formula or restrictions 
of it. This includes multi-solitons and soliton resonances (KP-II), and 
lump solutions (passing to KP-I via $t_{2n} \mapsto i \,  t_{2n}$ 
and performing suitable limits of parameters). 
\vskip.1cm

\noindent
{\bf Example 2.} If $M=N$ and $S=0$, let us consider
\be
  H = \mathcal{T} \, H_0 \, \mathcal{T}^{-1} \, , \qquad
  H_0 = \left( \begin{array}{cc} L & I \\ 0 & L \end{array} \right) 
         \, , \qquad
  \mathcal{T} = \left( \begin{array}{cc} I & -K \\ 0 & I \end{array} \right) \, ,
\ee
with $I = I_N$ and a constant $N \times N$ matrix $K$. As a consequence,
\be
          Q = I + [L,K] \; .
\ee 
We note that $H_0$ is \emph{not} similar to $\mathrm{diag}(L,L)$ \cite{Schw82}. 
Now we obtain
\be
    H^n = \mathcal{T} \, H_0^n \, \mathcal{T}^{-1}
        = \left(\begin{array}{cc} L^n & n \, L^{n-1} + [L^n , K] \\ 
              0 & L^n  \end{array}\right) 
\ee
and furthermore
\be
    e^{\xi(H)} = \left(\begin{array}{cc} e^{\xi(L)} & 
           \sum_{n \geq 1} n \, t_n \, L^{n-1} e^{\xi(L)} + [e^{\xi(L)} , K] \\ 
              0 & e^{\xi(L)}  \end{array}\right)  \; .
\ee
If $[L,[L,K]] \neq 0$ (which is condition (\ref{matrix_na_condition})), we 
obtain the solution 
\be
    \phi
 = e^{\xi(L)} \phi_0 \, (I + K \phi_0 + F )^{-1} e^{-\xi(L)} 
\ee
of the matrix pKP hierarchy in $(\mathcal{M}(N,N),\circ)$, where 
\be
  F := \Big( \sum_{n \geq 1} n \, t_n \, L^{n-1} 
       - e^{-\xi(L)} K e^{\xi(L)} \Big) \phi_0 \; .
\ee
Furthermore, using $F_{t_1} = e^{-\xi(L)} (I + [L,K]) e^{\xi(L)} \phi_0$, 
we find
\be
     \varphi 
  = \tr ( (I + [L,K]) \phi )
  =  \tr( F_{t_1} (I + K \phi_0 + F)^{-1} ) 
  = (\tr \log( I + K \phi_0 + F ))_{t_1}  
\ee
and thus
\be
   \varphi = (\log \tau)_{t_1}  \, , \qquad \quad
     \tau := \det\Big( I + K \phi_0 + ( \sum_{n \geq 1} n \, t_n \, L^{n-1} 
             - e^{-\xi(L)} K e^{\xi(L)} ) \phi_0 \Big) \; .
\ee
If $\mathrm{rank}(I + [L,K])=1$ (see also \cite{Shiota94,Wils98CM,Gekh+Kasm06} 
for appearances of this condition), then $\varphi$ solves the scalar pKP hierarchy. 
Assuming that $\phi_0$ is invertible, we can rewrite $\tau$ as follows,
\be
  \tau = \det\Big( e^{\xi(L)} (\phi_0^{-1} + K) e^{-\xi(L)} 
    + \sum_{n \geq 1} n \, t_n \, L^{n-1} - K \Big)   \label{ex2_tau}
\ee
(dropping a factor $\det(\phi_0)$). 
This simplifies considerably if we set $\phi_0^{-1} = - K$.\footnote{Note that 
in this case $\phi = (\sum_{n \geq 1} n \, t_n \, L^{n-1} - K )^{-1}$,  
which is rational in any finite number of the variables $t_n$. }
Choosing moreover 
\be
    L_{ij} = - (q_i - q_j)^{-1} \quad i \neq j \, , \quad
    L_{ii} = - p_i \, , \quad
    K = \mathrm{diag}(q_1, \ldots, q_N) \, ,
\ee
(\ref{ex2_tau}) reproduces a polynomial (in any finite number of the $t_n$) tau 
function associated with Calogero-Moser systems \cite{Shiota94,Wils98CM,Gekh+Kasm06}. 
Alternatively, we may choose
\be
    L = \mathrm{diag}(q_1, \ldots, q_N) \, , \quad
    K_{ij} = (q_i - q_j)^{-1} \quad i \neq j \, , \quad
    K_{ii} = p_i \; .
\ee
The corresponding solutions of the KP-I hierarchy ($t_{2n} \mapsto i \, t_{2n}$) 
include the rational soliton solutions (`lumps') originally obtained in 
\cite{MZBIM77}. In particular, $N=2$ and $q_2 = - \bar{q}_1$, 
$p_2 = \bar{p}_1$ (where the bar means complex conjugation), yields the 
single lump solution given by
\be
    \tau = | p_1 + \xi'(q_1)|^2 + \frac{1}{4 \Re(q_1)^2} \quad
    \mbox{where} \quad 
    \xi'(q) := \sum_{n \geq 1} n \, t_n \, q^{n-1} 
               \Big|_{ \{ t_{2k} \mapsto i \, t_{2k} \, , \, k=1,2,\ldots \}} \; .
\ee
\vskip.1cm

\noindent
{\bf Example 3.} Let $M=N$ and 
$L = S \pi_-$, $\, R = \pi_+ S$, $\, Q = \pi_+ S \pi_-$,  
with constant $N \times N$ matrices $\pi_+,\pi_-$ subject to 
$\pi_+ + \pi_- = I$. The matrix $H$ can then be written as
\be
     H = \left(\begin{array}{cc} \pi_+ \\ I \end{array}\right) 
         S \left(\begin{array}{cc} I & \pi_- \end{array}\right) \, ,
\ee
which lets us easily calculate 
\be
    H^n = \left(\begin{array}{cc} \pi_+ S^n & \pi_+ S^n \pi_- \\ 
              S^n & S^n \pi_- \end{array}\right) \; .
\ee
As a consequence, we obtain
\be
    \phi = (-C_+ + e^{\xi(S)} C_-)(\pi_{-} C_+ + \pi_{+} e^{\xi(S)} C_-)^{-1} \, ,
\ee
where $C_{\pm} := I \mp \pi_{\pm} \phi_0$. This solves the matrix pKP 
hierarchy in $\mathcal{M}(M,N)$ with the product $A \circ B = A \pi_+ S \pi_- B$ 
if (\ref{matrix_na_condition}) holds, which is 
$\pi_+ S (\pi_+ - \pi_-) S \pi_- \neq 0$. 
If furthermore $\mathrm{rank}(\pi_{+} S \pi_{-}) = 1$, then 
\be
    \varphi = \tr(Q \phi) = -\tr(\pi_{+}S) + (\log\tau)_{t_1} \, , \qquad
    \tau = \det(\pi_{-} C_+ + \pi_{+} e^{\xi(S)} C_-) 
\ee
solves the scalar pKP hierarchy (see also \cite{DMH06Burgers}). 
We will meet the basic structure underlying this example again in 
section~\ref{section:GD}. 
\vskip.1cm

\noindent
\emph{Remark~3.} A $GL(N+M,\mathbb{C})$ matrix
\be
    \mathcal{T} = \left( \begin{array}{cc} A & B \\ C & D \end{array} \right) 
\ee
can be decomposed as follows,
\be
 \mathcal{T} = \left( \begin{array}{cc} I_N & B D^{-1} \\ 0 & I_M \end{array} \right)
     \left( \begin{array}{cc} S_D & 0 \\ 0 & D \end{array} \right) 
     \left( \begin{array}{cc} I_N & 0 \\ D^{-1} C & I_M \end{array} \right) \, ,
\ee
if $D$ and its Schur complement $S_D = A - B D^{-1} C$ are both invertible. 
Let us see what effect the three parts of $\mathcal{T}$ induce on $\phi$ 
when acting on $Z$. \\
(1) Writing $P = D^{-1} C$, the first transformation leads to 
$\phi \mapsto \phi + P$, a shift by the constant matrix $P$. \\
(2) The second transformation amounts to $\phi \mapsto D \phi S_D^{-1}$ 
(where $\phi$ is now the result of the previous transformation). \\
(3) Writing $K = -B D^{-1}$, the last transformation is 
$\phi \mapsto \phi \, (I_N - K \phi)^{-1}$. 
\hfill $\square$

\section{WNA algebras and solutions of the discrete KP hierarchy}
\label{section:DKP}
\setcounter{equation}{0}
The potential discrete KP (pDKP) hierarchy in an associative algebra $(\mathcal{A},\circ)$ can be expressed in functional form as 
follows,\footnote{This functional 
representation of the pDKP hierarchy is equivalent to (3.32) in \cite{DMH06func}.}
\be
    \Omega(\la)^+ - \Omega(\la)_{-[\mu]}
  = \Omega(\mu)^+ -  \Omega(\mu)_{-[\la]} \, ,  \label{pDKP}
\ee
where $\la,\mu$ are indeterminates,
\be
    \Omega(\la) := \la^{-1} (\phi - \phi_{-[\la]}) 
                   - (\phi^+ - \phi_{-[\la]}) \circ \phi  \, ,
                  \label{pDKP_Omega}
\ee
and $\phi = (\phi_k)_{k \in \mathbb{Z}}$, $\phi_k^+ := \phi_{k + 1}$. 
The pDKP hierarchy implies that each component $\phi_k$, $k \in \mathbb{Z}$, 
satisfies the pKP hierarchy and its remaining content is a special pKP B\"acklund
transformation (BT) acting between neighbouring sites on the linear lattice 
labeled by $k$ \cite{Adle+vanM99,DMH06func}. 
This suggests a way to extend the method of section~\ref{section:matrix} 
to construct exact solutions of the pDKP hierarchy. 
What is needed is a suitable extension of (\ref{na_hier}) that accounts 
for the BT and this is offered by proposition~\ref{prop:naBT}.

\begin{theorem}
\label{theorem:DKP}
Let $\A$ be a WNA algebra with a constant element $\nu \in \A$, $\nu \not\in \A'$. 
Any solution 
\be
    f = (\nu - \phi_k)_{k \in \mathbb{Z}} \, ,
\ee
of the hierarchy (\ref{na_hier}) together with the compatible 
constraint\footnote{ Note that (\ref{pDKP_BT}) implies
$f^{n+} \ci_n f = 0$, where $f^{n+}_k := f_{k+n}$. This follows by induction from 
$f^{(n+1)+} \ci_{n+1} f = f^{(n+1)+} \circ (f^{n+} \ci_n f) - (f^{(n+1)+} \circ f^{n+}) \ci_n f = f^{(n+1)+} \circ (f^{n+} \ci_n f) - (f^+ \circ f)^{n+} \ci_n f$, 
where we used (\ref{circ_n}) and $[f^{n+}]=[f]$ in the first step.}
\be
    f^+ \circ f = 0   \label{pDKP_BT}
\ee
yields a solution $\phi = (\phi_k)_{k \in \mathbb{Z}}$ of the pDKP hierarchy 
in $\A'$. 
\end{theorem}
{\em Proof:} Since $[f^+]=[f]$, the compatibility follows by setting 
$f' = f^+$ and $\alpha=0$ in proposition~\ref{prop:naBT}. 
Using $f_{t_1} = f \circ f$, we rewrite (\ref{na_hier_func}) as
\bez
   \la^{-1} ( f - f_{-[\la]} ) + (f - f_{-[\la]}) \circ f - f_{t_1} = 0 \; .
\eez
Inserting $f = \nu - \phi$, this takes the form 
\bez
    \la^{-1} (\phi - \phi_{-[\la]}) - \phi_{t_1} 
  - (\phi - \phi_{-[\la]}) \circ \phi = \theta - \theta_{-[\la]}
\eez
with $\theta := - \phi \circ \nu$. Next we use (\ref{pDKP_BT}) and 
$f_{t_1} = f \circ f$ to obtain $(f^+ - f) \circ f + f_{t_1} = 0$, which is
\bez
   \phi_{t_1}  - (\phi^+ - \phi) \circ \phi = \theta^+ - \theta \;.
\eez
Together with the previous equation, this leads to
\bez
      \la^{-1} (\phi - \phi_{-[\la]}) 
   - (\phi^+ - \phi_{-[\la]}) \circ \phi = \theta^+ - \theta_{-[\la]} 
\eez
(which is actually equivalent to the last two equations), so that
\bez
     \Omega(\la) = \theta^+ - \theta_{-[\la]} \; . 
\eez
This is easily seen to solve (\ref{pDKP}). 
\hfill $\square$
\vskip.1cm

Let us choose the WNA algebra of section~\ref{section:matrix}.\footnote{Since  
there is only a single element $\nu$, the matrices $L,R,S$ do not depend 
on the discrete variable $k$.}
Evaluation of (\ref{na_hier}) leads to the matrix Riccati hierarchy 
(\ref{matrix_Riccati}), and (\ref{pDKP_BT}) with $f^+ = \nu + C - \phi^+$ 
becomes
\be
   S + C R + (L + C Q) \phi - \phi^+ R - \phi^+ Q \phi = 0 \, ,  \label{pDKP_RiccatiBT}
\ee
which can be rewritten as 
\be
    \phi^+ 
   = (S + L \phi)(R + Q \phi)^{-1} + C 
   = Y^+ \, (X^+)^{-1}   \label{DKP_phi+}
\ee
(assuming that the inverse matrices exist), where $X^+, Y^+$ are the 
components of
\be
    Z^+ = T H \, Z = T H \, e^{\xi(H)} Z^{(0)}  \, ,   \label{pDKP_Z+}
\ee
with $Z, H, T$ taken from section~\ref{section:matrix}. Deviating from
the notation of section~\ref{section:matrix}, we write $Z^{(0)}$ for the 
constant vector, since $Z_0$ should now denote the component of $Z$ 
at the lattice site $0$. In order that (\ref{pDKP_Z+}) 
defines a pDKP solution on the whole lattice, we need $H$ invertible. 
Since the matrix $C$, and thus also $T$, may depend on the lattice site $k$, 
solutions of (\ref{pDKP}) are determined by 
\be
    Z_k = T_k H T_{k-1} H \cdots T_1 H Z_0 \, , \quad
    Z_{-k} = (T_{-k} H)^{-1} (T_{-k+1} H)^{-1} \cdots (T_{-1} H)^{-1} Z_0 
    \, , \qquad  k \in \mathbb{N} \; .
\ee
This corresponds to a sequence of transformations applied to 
the matrix pKP solution $\phi_0$ determined by $Z_0$, which generate 
new pKP solutions (cf. \cite{Adle+vanM99}). 
$\phi_1$ is then given by (\ref{DKP_phi+}) in terms of $\phi_0$, and 
\be
   \phi_2 &=& [ L S + S R + L C_1 R + ( L^2 + S Q + L C_1Q ) \phi_0 ] 
              \nonumber  \\
          & & \times
   [ R^2 + Q S + Q C_1 R + ( QL + R Q + Q C_1 Q ) \phi_0 ]^{-1} + C_2 
\ee
shows that the action of the $T_k$ becomes considerably more involved for $k>1$.  
In the special case $T_k = I_{N+M}$ (so that $C_k =0$), we have
\be
    Z_k = e^{\xi(H)} (H^k Z^{(0)}_0)  \qquad  k \in \mathbb{Z} \; .
\ee
If $X^{(0)}_k, Y^{(0)}_k$ are the components of the vector $H^k Z^{(0)}_0$, 
the lattice component $\phi_k$ of the pDKP solution determined in this way 
is therefore just given by the pKP solution of section~\ref{section:matrix} 
with initial data (at $\mathbf{t}=0$) 
\be
    \phi^{(0)}_k 
  = Y^{(0)}_k (X^{(0)}_k)^{-1}
  = L^k \phi^{(0)}_0 \, [R^k + (R^k K - K L^k) \phi^{(0)}_0 ]^{-1} 
    \; .
\ee
With the restrictions of example~1 in section~\ref{section:matrix}, 
assuming that $L$ and $R$ are invertible (so that $H$ is invertible), 
the corresponding solution of the matrix pDKP hierarchy (in the 
matrix algebra with product $A \circ B = A(RK-KL)B$) is
\be
    \phi_k = e^{\xi(L)} L^k \phi^{(0)}_0 [ R^k (I_N + K \phi^{(0)}_0) 
             - e^{-\xi(R)} K e^{\xi(L)} L^k \phi^{(0)}_0 ]^{-1} e^{-\xi(R)} \, ,
    \qquad\quad  k \in \mathbb{Z} \, , 
\ee
which leads to
\be
   \varphi_k = (\log \tau_k)_{t_1}  
   \qquad \mbox{with} \quad 
    \tau_k = \det\left( R^k (I_N + K \phi^{(0)}_0 ) 
             - e^{-\xi(R)} K e^{\xi(L)} L^k \phi^{(0)}_0 \right)
    \qquad k \in \mathbb{Z} \; .
\ee
If $Q = RK-KL$ has rank 1, this is a solution of the scalar pDKP 
hierarchy.\footnote{Recall that $\varphi = \tr(Q \phi)$ (cf. \ref{varphi}) 
determines a homomorphism if $Q$ has rank 1. As a consequence, if $\phi$ 
solves the matrix pDKP hierarchy (\ref{pDKP}), then $\varphi$ solves 
the scalar pDKP hierarchy.} 
As a special case, let us choose $M=N$, $L = \mathrm{diag}(p_1,\ldots,p_N)$, 
$R=\mathrm{diag}(q_1,\ldots,q_N)$, and $K$ with entries 
$K_{ij} = (q_i-p_j)^{-1}$.\footnote{The condition (\ref{matrix_na_condition}) 
requires $q_i \neq p_j$ for all $i,j=1,\ldots,N$.}
 Then $Q$ has rank 1 and we obtain  
$N$-soliton tau functions of the scalar discrete KP hierarchy. 
These can also be obtained via the Birkhoff decomposition 
method using appropriate initial data as in  \cite{Ueno+Taka84,Saka04}. 

With the assumptions made in example~2 of section~\ref{section:matrix}, 
setting $\phi^{(0)}_0 = -K^{-1}$, assuming that $K$ and $L$ are invertible, 
and choosing for $T_k$ the identity, we find the matrix pDKP solution
\be
   \phi_k = \Big( \sum_{n \geq 1} n \, t_n \, L^{n-1} + k \, L^{-1} - K \Big)^{-1} 
   \, , \qquad\quad  k \in \mathbb{Z} \; . 
\ee
If $\mathrm{rank}(I_N + [L,K])=1$, this leads to the following solution of the 
scalar pDKP hierarchy, 
\be
   \varphi_k = (\log \tau_k)_{t_1}  \qquad \mbox{with} \quad
   \tau_k = \det \Big( \sum_{n \geq 1} n \, t_n \, L^{n-1} + k \, L^{-1} - K \Big) \; .
\ee

In example~3 of section~\ref{section:matrix}, $H$ is not invertible, 
so that (\ref{pDKP_Z+}) does not determine a pDKP solution.

\section{From WNA to Gelfand-Dickey}
\label{section:GD}
\setcounter{equation}{0}
Let $\R$ be the complex algebra of pseudo-differential operators \cite{Dick03}
\be 
	\mathcal{V} = \sum_{i \ll \infty} v_i \, \pa^i \, ,
\ee 
with coefficients $v_i \in \fA$, where $\fA$ is the complex differential
algebra of polynomials in (in general noncommuting) symbols $u_n^{(m)}$, 
$m=0,1,2,\ldots$, $n=2,3,\ldots$, where $\pa(u_n^{(m)}) = u_n^{(m+1)}$ 
and $\pa(vw) = \pa(v) \, w + v \, \pa(w)$ for $v,w \in \fA$. 
We demand that $u_n^{(m)}$, $n=2,3,\ldots$, $m=0,1,2,\ldots$, 
are algebraically independent in $\fA$, and 
we introduce the following linear operators on $\R$, 
\be
    S(\mathcal{V}) := \mathfrak{L} \mathcal{V} \, , \qquad 
    \pi_+(\mathcal{V}) := \mathcal{V}_{\geq 0} \, , \qquad
    \pi_-(\mathcal{V}) := \mathcal{V}_{<0} := \mathcal{V} - \mathcal{V}_{\geq 0} \, ,
\ee
where $\mathcal{V}_{\geq 0}$ is the projection of a pseudo-differential 
operator $\mathcal{V}$ to its differential operator part, and 
\be
    \mathfrak{L} = \pa + u_2 \, \pa^{-1} + u_3 \, \pa^{-2} + \cdots \; .
\ee

Let $I$ denote the identity of $\R$ (which we identify with the identity 
in $\fA$), and let $\mathcal{O}$ be the subspace of linear operators 
on $\R$ spanned by $S$ and elements of the form
$S \pi_\pm S \pi_\pm \cdots \pi_\pm S$ 
(with any combination of signs). $\mathcal{O}$ becomes an algebra 
with the product given by
\be
	A \circ B := A \pi_+ S \pi_- B  \; .
\ee
$(\mathcal{O}, \circ)$ is then generated by the elements
$(S \pi_-)^m S \, (\pi_+ S)^n$, $m,n=0,1,\ldots$.
Let us furthermore introduce 
$\cA := \{v \in \fA \, : \, v = \res(A(I)), \; A \in \mathcal{O} \}$, 
where $\mathrm{res}$ takes the residue (the coefficient of $\pa^{-1}$) 
of a pseudo-differential operator. 
This is a subalgebra of $\fA$, since for $A,B \in \mathcal{O}$ we have
\be	
	\res(A(I)) \, \res(B(I)) = \res(A \pi_+ S \pi_- B(I)) \, , 
		\label{res_homomorphism}
\ee
so that the product of elements of $\cA$ is again in $\cA$. As a consequence 
of this relation (read from right to left), $\cA$ is generated by the 
elements $\res((S\pi_-)^m S (\pi_+ S)^n(I))$, $m,n=0,1,\ldots$.  
Based on the following preparations, we will argue that $\cA$ and $(\mathcal{O},\ci)$ are actually isomorphic algebras.

\begin{lemma}
For all $\mathcal{V} \in \R$, 
\be
     \res( (S \pi_-)^m \mathcal{V} ) 
   = \res( \mathcal{D}_m \, \mathcal{V} ) \, , \qquad m=0,1,\ldots \, ,
\ee
where $\mathcal{D}_0 = I$ and $\{\mathcal{D}_m \}_{m=1}^\infty$ are 
the differential operators recursively determined
by $\mathcal{D}_m = (\mathcal{D}_{m-1} \mathfrak{L})_{\geq 0}$. 
\end{lemma}
{\em Proof:} We do the calculation for $m=2$. This is easily generalized to 
arbitrary $m \in \mathbb{N}$. 
\bez
      \res( (S \pi_-)^2 \, \mathcal{V} ) 
   = \res( \mathfrak{L} (\mathfrak{L} \mathcal{V}_{<0} )_{<0} ) 
   = \res( \mathfrak{L}_{\geq 0} \mathfrak{L} \mathcal{V}_{<0} ) 
   = \res( (\mathfrak{L}_{\geq 0} \mathfrak{L})_{\geq 0} \mathcal{V} ) 
   = \res( \mathcal{D}_2 \, \mathcal{V} ) \; . 
\eez
\begin{minipage}{16cm}
\vspace{-1.5cm}
\hfill $\square$
\end{minipage}
\vspace{-1.2cm}

\begin{proposition}
\be
   \res( (S \pi_-)^m S (\pi_+S)^n(I) ) 
 = \sum_{k=0}^m {m \choose k} u^{(k)}_{m+n+2-k} 
   + \mbox{terms nonlinear in $u_k^{(j)}$} \, ,  \quad m,n=0,1,\ldots
\ee
\end{proposition}
{\em Proof:} According to the preceding lemma, we have
\bez
   \res( (S \pi_-)^m S (\pi_+S)^n(I) ) 
 = \res( \mathcal{D}_m S (\pi_+S)^n(I) ) \; .
\eez
Next we note that $\mathcal{D}_m = \pa^m + D_m$,   
$(\pi_+ S)^n(I) = \pa^n + D'_n$ with differential operators $D_m, D'_n$ 
(of degree smaller than $m$, respectively $n$) such that each of its summands 
contains factors from $\{ u_k^{(j)} \}$ (so their coefficients are 
non-constant polynomials in the $u_k^{(j)}$). It follows that
\bez
     \res( (S \pi_-)^m S (\pi_+S)^n(I) ) 
 &=& \res( (\pa^m + D_m) \mathfrak{L}_{<0} (\pa^n + D'_n) )  \\
 &=& \res( \pa^m \mathfrak{L}_{<0} \pa^n ) + \mbox{terms nonlinear in $u_k^{(j)}$} \; .
\eez
It remains to evaluate 
\bez
     \res( \pa^m \mathfrak{L}_{<0} \pa^n ) 
 &=& \sum_{j=1}^\infty \res( \pa^m u_{1+j} \pa^{n-j}) 
  = \sum_{j=1}^\infty \res\Big( \sum_{k=0}^m {m \choose k} 
     u^{(k)}_{1+j} \pa^{m+n-j-k} \Big) \\
 &=& \sum_{k=0}^m {m \choose k} u^{(k)}_{m+n+2-k} \; .
\eez
\begin{minipage}{16cm}
\vspace{-1.5cm}
\hfill $\square$
\end{minipage}
\vspace{-.7cm}

According to the last proposition, the linear term with the highest 
 derivative\footnote{If $m=0$, the linear term is simply $u_{n+2}$ and 
thus again `the linear term with the highest derivative'.} 
in the residue of $(S \pi_-)^m S (\pi_+ S)^n(I)$ is given by $u_{n+2}^{(m)}$. 
We conclude that the monomials $(S \pi_-)^m S (\pi_+ S)^n$, 
$m,n=0,1,\ldots$, are algebraically independent in $(\mathcal{O},\ci)$, 
since any algebraic relation among them 
would induce a corresponding algebraic relation in the set of $u_n^{(m)}$, 
but we assumed the $u_n^{(m)}$ to be algebraically independent. 
Together with (\ref{res_homomorphism}), this implies that 
$\cA$ and $(\mathcal{O},\ci)$ are isomorphic algebras. 

The last result allows us to introduce a WNA structure directly on $\cA$ 
as follows.\footnote{Note that the corresponding WNA structure for 
$(\mathcal{O},\ci)$ resembles that of example~3 in section~\ref{section:matrix}.}
Augmenting $\cA$ with $f$ such that, for 
$\mathcal{V} ,\mathcal{W} \in \mathcal{O}(I)$, 
\be
  f \circ f := -\res(\mathfrak{L}) \, , &\qquad&
  f \circ \res(\mathcal{V}) := \res(\mathfrak{L} \mathcal{V}_{<0}) \, , \nonumber \\
  \res(\mathcal{V}) \circ f := -\res(\mathcal{V}_{<0} \mathfrak{L}) \, , &\qquad&
    \res(\mathcal{V}) \circ \res(\mathcal{W}) 
 := \res(\mathcal{V}) \, \res(\mathcal{W}) \, ,  \label{res_WNA_rules}
\ee 
indeed defines a WNA algebra $\A = \A(f)$.
The relations (\ref{res_WNA_rules}) are well-defined since $\res(A(I))$ 
uniquely determines $A \in \mathcal{O}$. By induction we obtain
\be 
  f \ci_n f = - \res(\mathfrak{L}^n) \, , &\qquad& 
  f \ci_n \res(\mathcal{V}) = \res(\mathfrak{L}^n \mathcal{V}_{<0}) \, , \nonumber \\
 \res(\mathcal{V}) \ci_n f = - \res(\mathcal{V}_{<0} \mathfrak{L}^n) \, , && 
   \res(\mathcal{V}) \ci_n \res(\mathcal{W}) 
 = \res(\mathcal{V}_{<0} \mathfrak{L}^n \mathcal{W}_{<0}) \; .
\ee   

Let the $u_n$ now depend on variables $t_1,t_2,\ldots$, 
and set $\pa = \pa_{t_1}$. The hierarchy (\ref{na_hier}) of ODEs, 
\be
    f_{t_n} = f \ci_n f = -\res(\mathfrak{L}^n) \, ,  
    \qquad \quad   n=1,2,\ldots \, ,  \label{ps_ode_hier_S}
\ee 
by use of the WNA structure implies
\be
    \pa_{t_n}(\res(\mathfrak{L}^m))
 &=& -\pa_{t_n}(f \ci_m f)
  = - f_{t_n} \ci_m f - f \ci_m f_{t_n} \nonumber \\
 &=& -(f \ci_n f) \ci_m f - f \ci_m (f \ci_n f) 
  = \res\Big( \mathfrak{L}^m (\mathfrak{L}^n)_{<0} - (\mathfrak{L}^n)_{\geq 0} 
    \mathfrak{L}^m \Big) \nonumber \\
 &=& \res\Big( [(\mathfrak{L}^n)_{\geq 0},\mathfrak{L}^m] \Big) \; .   \label{res(GD)}
\ee
Since also 
$\pa_{t_n}(\mathrm{res}(\mathfrak{L}^m)) = \mathrm{res}( [(\mathfrak{L}^m)_{\geq 0} , \mathfrak{L}^n] ) = \pa_{t_m}(\mathrm{res}(\mathfrak{L}^n))$, 
we conclude that if we extend $\cA$ to $\tilde{\cA}$ by adjoining an element 
$\phi =\pa^{-1}(u_2)$, then
\be
    \phi_{t_n} = \mathrm{res}(\mathfrak{L}^n) \, , 
    \qquad \quad   n=1,2,\ldots   \; .  \label{GD_phi_tn}
\ee
It follows that $\nu := f+\phi$ satisfies $\pa_{t_n}(\nu)=0$, $n=1,2,\ldots$, 
and is therefore constant. (\ref{GD_phi_tn}) determines all the $u_k$ in terms 
of the derivatives of $\phi$ (see \cite{DMH04ncKP}, for example). 
 From (\ref{GD_phi_tn}) with $n=2,3$, and (\ref{res(GD)}) with $m=n=2$,  
we recover the pKP equation
\be
    (4 \phi_{t_3} - \phi_{t_1 t_1 t_1} - 6 \, \phi_{t_1}{}^2)_{t_1}
  - 3 \, \phi_{t_2 t_2} + 6 \, [\phi_{t_1},\phi_{t_2}] = 0 \, ,
\ee
in accordance with the general theory. More generally, the equations  
(\ref{res(GD)}) determine the whole pKP hierarchy. They are the residues of 
\be
  \pa_{t_n}(\mathfrak{L}^m) = [(\mathfrak{L}^n)_{\geq 0} , \mathfrak{L}^m] \, , 
  \qquad \quad   m,n =1,2, \ldots \; . 
\ee
This is equivalent to the Gelfand-Dickey (GD) system
$\pa_{t_n}(\mathfrak{L}) = [(\mathfrak{L}^n)_{\geq 0} , \mathfrak{L}]$,  
$n=1,2,\ldots$, which is a well-known formulation of the KP hierarchy 
(see \cite{Dick03}, for example).

We have thus shown how the Gelfand-Dickey formulation of the KP hierarchy 
can be recovered in the WNA framework. In fact, for the particular WNA algebra 
chosen above, the hierarchy (\ref{na_hier}) of ODEs is equivalent to the 
Gelfand-Dickey formulation of the KP hierarchy.

\section{Conclusions}
\label{section:concl}
\setcounter{equation}{0}
In this work we extended our previous results \cite{DMH06nahier,DMH06CJP} 
on the relation between weakly nonassociative (WNA) algebras and solutions 
of KP hierarchies to discrete KP hierarchies. We also provided further examples 
of solutions of matrix KP hierarchies and corresponding solutions of the scalar 
KP hierarchy. 
In particular we recovered a well-known tau function related to 
Calogero-Moser systems in this way (example~2 in section~\ref{section:matrix}). 
Furthermore, we established a connection with the Gelfand-Dickey formulation 
of the KP hierarchy. 
As a byproduct, in section~\ref{section:GD} we obtained a new realization 
of the \emph{free} WNA algebra generated by a single element, which 
also has a realization in terms of quasi-symmetric functions 
\cite{DMH06nahier}. 
There is more, however, we have to understand in the WNA framework. 
In particular this concerns the multi-component KP hierarchy (see 
\cite{Kac+vanderLeur03} and references therein) and its 
reductions, which include the 
Davey-Stewartson, two-dimensional Toda lattice and $N$-wave hierarchies. 
Our hope is that also in these cases the WNA approach leads in a quick way 
to relevant classes of exact solutions.

\section*{Appendix: From Riccati to KP with FORM}
The following FORM program \cite{Heck00,Verm02} verifies that any solution 
of the first three equations of the Riccati hierarchy (\ref{matrix_Riccati}) 
solves the pKP equation in an algebra with product $A \circ B = AQB$. 
\vspace{-.1cm}
\begin{verbatim}
Functions phi,phix,phiy,phit,L,Q,R,S,dx,dy,dt;  Symbol n;
Local pKP = dx*(4*phit - 6*phix*Q*phix - dx^2*phix) - 3*dy*phiy 
 + 6*( phix*Q*phiy - phiy*Q*phix );       * pKP equation
repeat;
id phix = S + L*phi - phi*R - phi*Q*phi;  * Riccati system
id phiy = S(2) + L(2)*phi - phi*R(2) - phi*Q(2)*phi;
id phit = S(3) + L(3)*phi - phi*R(3) - phi*Q(3)*phi;
id dx*phi = phix + phi*dx;   * product rule of differentiation
id dy*phi = phiy + phi*dy;  id dt*phi = phit + phi*dt;
id dx?{dx,dy,dt}*L?{L,Q,R,S} = L*dx;      * L,Q,R,S are constant
* recursion relations for matrices (see proof of proposition 2):
id L(n?{2,3}) = L*L(n-1) + S*Q(n-1); 
id R(n?{2,3}) = Q*S(n-1) + R*R(n-1); 
id S(n?{2,3}) = L*S(n-1) + S*R(n-1); 
id Q(n?{2,3}) = Q*L(n-1) + R*Q(n-1);
id L?{L,Q,R,S}(1) = L;  
endrepeat;
id dx?{dx,dy,dt} = 0;  
print pKP;         * should return zero
.end
\end{verbatim}
\vspace{-.1cm}
This program provides an elementary and quick way toward the classes 
of exact solutions of the KP equation given in the examples in 
section~\ref{section:matrix}.

\end{document}